%% file: main.tex
\documentclass[aps,twocolumn,preprintnumbers,superscriptaddress]{revtex4}

\usepackage{graphicx}  
\usepackage{subfigure}
\usepackage{multirow}

\usepackage{hyperref} 

\usepackage{fancyhdr}
\usepackage{longtable}
\usepackage{xcolor}
\usepackage{parskip}
\usepackage[T1]{fontenc}
\usepackage{dcolumn}   

\usepackage{bm}        
\usepackage{amsfonts}  
\usepackage{amsmath}   
\usepackage{amssymb}   
\usepackage{ulem}      
\usepackage{slashed}
\usepackage{braket}

\usepackage[T1]{fontenc}
\usepackage[utf8]{inputenc}
\usepackage{lmodern}

\usepackage{tikz}
\tikzset{every picture/.style={line width=0.75pt}} 

\newcommand{\trace}[1]{\mathrm{Tr} \left( #1 \right)}

\newenvironment{envCom}
  {\color{cyan}%
   \textbf{\small Comment: }\ignorespaces}
 {\smallskip}
 
 \newenvironment{envAdd}
  {\color{blue}%
   \textbf{\tiny [Add] }\ignorespaces}
 {\smallskip}
 
 \newenvironment{envRem}
  {\color{red}\ignorespaces}
 {\smallskip}

\begin{document}

\title{Controlled State Reconstruction and Quantum Secret Sharing}


\author{Pahulpreet Singh}
\email{pahulpreet.singh@research.iiit.ac.in}
\affiliation {Centre for Quantum Science and Technology\\International Institute of Information Technology Hyderabad, Gachibowli, Hyderabad -- 500032, Telangana, India}
\affiliation {Center for Computational Natural Science and Bioinformatics\\International Institute of Information Technology Hyderabad, Gachibowli, Hyderabad -- 500032, Telangana, India}

\author{Indranil Chakrabarty}
\email{indranil.chakrabarty@iiit.ac.in}
\affiliation {Centre for Quantum Science and Technology\\International Institute of Information Technology Hyderabad, Gachibowli, Hyderabad -- 500032, Telangana, India}
\affiliation{Center for Security, Theory and Algorithmic Research\\International Institute of Information Technology Hyderabad, Gachibowli, Hyderabad -- 500032, Telangana, India}

\begin{abstract}
    In this article, we present a benchmark for resource characterization in the process of controlled quantum state reconstruction and secret sharing for general three-qubit states. This is achieved by providing a closed expression for the reconstruction fidelity, which relies on the genuine tripartite correlation and the bipartite channel between the dealer and the reconstructor characterized by the respective correlation parameters. We formulate the idea of quantum advantage in approximate state reconstruction as surpassing the classical limit set at 2/3. This article brings out new interoperability between teleportation and state reconstruction. This is detailed through a case-by-case analysis of relevant correlation matrices. We are reformulating the idea of quantum secret sharing by setting up additional constraints on the teleportation capacity of the bipartite channels between dealer and shareholders by ensuring that individually, the shareholders cannot reconstruct the secret. We believe that this will give us the ideal picture of how quantum secret sharing should be.
\end{abstract}
\maketitle

\section{Introduction}

With the advent of quantum information theory, sending and sharing quantum \cite{1, 2, 3, 4, 5, 6, 7} as well as classical \cite{8, 9, 10, 11, 12, 13} information using quantum resources has become a significant area of study. During the transmission, security of the message becomes a key consideration and has to be taken care of to set up quantum networks \cite{14, 15, 16, 17, 18, 19, 20, 21} in the longer run. Secret sharing is a procedure that allows secure distribution of a secret message into multiple ($n$) parts so that a certain number ($k$) of parts can be used to get back the original message, and forbids any subset smaller than $k$ from obtaining useful information about it. Quantum secret sharing (QSS) \cite{22, 23, 24, 25, 26, 27, 28, 29, 30, 31, 32, 33, 34, 35, 36} is the extension of this idea where we use a quantum resource (usually a multi-party entangled state) to distribute the secret \cite{26, 27, 28}. The secret itself can be classical or quantum in nature, and of any size. In this article, we are interested in sharing of a qubit with two parties such that when both the parties co-operate, they can reconstruct it at one of the locations. Here, we name this process as controlled state reconstruction (CSR). CSR is a slightly broader term than the secret sharing in the sense that it need not account for the security aspects that come with secret sharing. The security here is defined by the condition that no share holders can reveal the secret without the help of the other share holders. This process has also been termed as controlled quantum teleportation (CQT) since it is analogous to teleporting a qubit from an initial location (\textit{dealer}) to a final location (\textit{reconstructor}) with the help of a third party (\textit{assistant}) acting as the control qubit \cite{37, 38, 39}. Recently, it has been shown that a witness operator can be constructed to estimate the power of the controller in CQT \cite{40}. \\
\noindent In this paper, we differentiate between the two scenarios by using the term "quantum secret sharing" (QSS) specifically when security is a requirement (i.e. individual shareholders do not have enough information about the secret) and "controlled state reconstruction" (CSR) as the general term for transmitting quantum information from one location to another with the help of an assistant. Most of our work focuses on CSR, however at the end, we investigate some intricacies that come with QSS.\\
\noindent We consider the simple setting where a \textit{dealer} (say, Alice) aims to share an \textit{unknown} qubit with two shareholders (say, Bob and Charlie).
At the start of the protocol, one of the shareholders is decided to be the \textit{re-constructor} (Charlie, for example) which makes the other one (Bob, in this case) the \textit{assistant}. Prefect reconstruction of the state after sharing has already been shown using the GHZ state as the resource \cite{22}. In this paper, we study approximate controlled reconstruction of the state using general tripartite entangled states. Hence, if we are using a resource which does not allow perfect reconstruction, we want the final qubit to be as close to the original qubit as possible. By quantum advantage, we refer to a situation where the reconstruction fidelity is better than what can be achieved classically without having any shared quantum resource. \\
\noindent Our first finding is the classical limit of controlled reconstruction which happens to be $2/3$. This matches with the classical limit of teleportation fidelity of a bi partite two qubit channel \cite{2}. Subsequently, we find an expression for the reconstruction fidelity  in terms of the Bloch parameters of the resource state.  It then enables us to find conditions under which a three-qubit states will have quantum advantage in the process of CSR/CQT. It turns out that the reconstruction fidelity is dependent on two siginificant parameters -- one is the correlation tensor between three parties, and the other one is correlation matrix of the dealer - reconstructor pair. It is intriguing to note that this fidelity does not depend on the correlation matrix of the assistant - reconstructor pair. In a sense, this fidelity just quantifies how much quantum information from the initial state is transferred to the final location. There can be situations with quantum advantage despite the absence of tripartite correlation. This leads us to believe that this fidelity does not originate solely from CSR. Rather, there is also contribution of the teleportation capacity of the bipartite channel between the dealer and the reconstructor. We not only prove this but consider several cases based on the correlation matrices that appear in our expression. This helps us analyze the interoperability between the reconstruction fidelity and the teleportation fidelity for three-qubit resources in a holistic manner. In the later part of our article we develop this idea further to introduce QSS in a different but meaningful way. This paves the way to distinguish QSS from CSR/CQT. For a QSS protocol to be successful, we make sure that neither of the share holders have useful information about the secret by enforcing additional constraints on the teleportation capacities of the dealer to share holder's channels. We also specify the states that can be used as resources for successful QSS. \\
\noindent In section \ref{sec:limit}, we obtain the classical limit of this fidelity without  using any quantum channel to share the state. In section \ref{sec:advantage}, we give an expression for the maximum possible reconstruction fidelity using the parameters of the resource state. This lets us quantify the quantum advantage an arbitrary three-qubit resource can provide in  reconstruction. In the same section, we discuss a potential relationship between (2-party) quantum teleportation and (3-party) state reconstruction. At the end, in section \ref{sec:secret}, we give criteria to discern QSS from CSR by adding additional constraints that need to be satisfied for QSS.

\section{Classical Limit of Controlled state Reconstruction}
\label{sec:limit}

We define the classical limit as the expected fidelity score obtained if only classical channels are used to share a qubit. This will allow us to define the threshold above which a quantum advantage can be claimed.
Here, we consider a three-party scheme where Alice is the dealer, Bob is the assistant, and Charlie is the reconstructor.
Let $\ket{q}$ denote Alice's qubit which is to be shared. She can measure in some basis (say {$\ket{\uparrow}$ , $\ket{\downarrow}$}), which would agreed upon by the parties beforehand. Then this measurement result can be encoded into a single classical bit \textit{s} (say 0 for $\ket{\uparrow}$ and 1 for $\ket{\downarrow}$). This can be split into two shares $s_1$ and $s_2$ such that $s_1 \oplus s_2 = s$
It can be shown that (for a given \textit{s}), it would be optimal for Alice to choose the appropriate $s_1$, $s_2$ from a uniform distribution (see Appendix \ref{ss:dishonest}).
Alice transmits the respective bits to Bob and Charlie through classical channels. During the reconstruction phase, Bob and Charlie cooperate to find \textit{s}, using which they can construct the corresponding quantum state $\ket{s}$ (again, $\ket{\uparrow}$ for s=0 and $\ket{\downarrow}$ for s=1) which is an approximation of Alice's original quantum state $\ket{q}$, which we represent as:
\begin{equation}
\ket{q} = \cos\frac{\theta}{2}\ket{\uparrow} + e^{i\phi}\sin\frac{\theta}{2}\ket{\downarrow}    
\end{equation}
\noindent Since the final state $\ket{s}$ is eventually dependent only on Alice's measurement, we can say $\ket{s}$ is $\ket{\uparrow}$ with probability $\cos^2\frac{\theta}{2}$ and $\ket{\downarrow}$ with probability $\sin^2\frac{\theta}{2}$. Now, calculating the fidelity between the two qubit states, 
\begin{eqnarray}
     F(q, s) &&= \| \braket{q|s} \|^2 {}\nonumber\\&&
    = Pr[s = 0] \cdot \| \braket{q|\uparrow} \|^2 + Pr[s = 1] \cdot \| \braket{q|\downarrow} \|^2 {}\nonumber\\&&
    = 1 - \frac{1}{2} \sin^2 \theta .
\end{eqnarray}
\noindent Taking the expectation fidelity over all states $|q\rangle$ on the Bloch sphere,
\begin{eqnarray}
    \mathcal{F} &&= \langle F(q, s) \rangle {}\nonumber\\&&
    = \frac{1}{4\pi} \int_{\theta = 0}^\pi \int_{\phi = 0}^{2\pi} \sin\theta \,d\theta \,d\phi (1 - \frac{1}{2} \sin^2 \theta) {}\nonumber\\&&
    = \frac{1}{2} (2) - \frac{1}{4} (\frac{4}{3}) = \boxed{\frac{2}{3}} .
\end{eqnarray}

\noindent Hence the Fidelity for reconstruction is $\mathcal{F}_c = \dfrac{2}{3}$. \\
\\
\noindent This is the value of the classical limit of the reconstruction fidelity of the state shared by the dealer. If one is able to achieve a fidelity more than this with the help of a shared quantum resource (in this case, a tripartite entangled state), we say that there is a quantum advantage.\\

\noindent \textbf{Note:} \textit{We recently became aware of \cite{41} who have done this calculation for a similar scenario and arrived at the same value, i.e. 2/3.}
 

\section{Approximate Controlled Reconstruction And Quantum Advantage}
\label{sec:advantage}

We start with a three-qubit resource state $\rho_{ABC}$ in the space $\mathcal{H}_A \otimes \mathcal{H}_B \otimes \mathcal{H}_C$. We can write it in parametric form as

\begin{eqnarray}
 &&\rho_{ABC}= 
 \frac{1}{8}[I^{\otimes 3} +\sum_{i=1}^3 a_i.\sigma _i\otimes I^{\otimes 2}+ {}\nonumber\\&&
 \sum_{j=1}^3 I \otimes b_j.\sigma_j \otimes I + \sum_{k=1}^3 I^{\otimes 2} \otimes c_k.\sigma_k+ {}\nonumber\\&&
 \sum_{i,j=1}^3 q_{ij} \sigma_i \otimes \sigma_j \otimes I + \sum_{i,k=1}^3 r_{ik}\sigma_i \otimes I \otimes \sigma_k + {}\nonumber\\&&\sum_{j,k=1}^3 s_{jk} I\otimes \sigma_j \otimes \sigma_k +\sum_{i,j,k=1}^3 t_{ijk} \sigma_i \otimes \sigma_j \otimes \sigma_k
 ].   \label{eq:resource_state}
\end{eqnarray}

\noindent Here $a_i,b_j,c_k$ are local Bloch vectors and the correlation matrices are given by,
$Q=\{q_{ij}\}=Tr(\rho_{ABC}(\sigma_i \otimes \sigma_j \otimes I)), R=\{r_{ik}\}=Tr(\rho_{ABC}( \sigma_i \otimes I \otimes \sigma_k))$ and $S=\{s_{jk}\}=Tr(\rho_{ABC}( I \otimes \sigma_j \otimes \sigma_k))$ are the correlation matrices of order $3 \times 3$. Here $\tau=t_{ijk}=Tr(\rho_{ABC}( \sigma_i \otimes \sigma_j \otimes \sigma_k))$ is the correlation tensor.

\noindent Now we find the  reconstruction fidelity of the state shared in terms of the Bloch parameters of the three-qubit resource state $\rho_{ABC}$. The  qubit $\S$ on Alice's side, parameterized by the Bloch vector {\boldmath$\phi$}, is given by
\begin{equation}
 \rho_{\S}=\frac{1}{2}(I+\sum_i\phi_i.\sigma_i).   \label{eq:secret}
\end{equation}

\noindent In the standard  scheme, measurement takes place at two phases of the protocol. First, at Alice's side on the to be shared qubit ($\S$) along with Alice's share of the resource ($A$), with projectors $P_l=|\Psi_l\rangle \langle \Psi_l|$ ($l=0,1,2,3$) and second, on Bob's qubit, with projectors $P_x=|x\rangle \langle x|$ ($x=+,-$). Here, the Bell states and Hadamard states are given as, $|\Psi_{(0)}^{3}\rangle=\frac{1}{\sqrt{2}}(|01\rangle \pm |10\rangle), |\Psi_{(1)}^{2}\rangle=\frac{1}{\sqrt{2}}(|00\rangle \pm |11\rangle)$ and $|x_{\pm}\rangle=\frac{1}{\sqrt{2}}(|0\rangle \pm |1\rangle)$ respectively. The Bell state projectors can be written in the form
\begin{equation}
    P_l = \frac{1}{4}(I^{\otimes 2} +  \sum_{ij} t_{ij} \sigma_i \otimes \sigma_j) .
\end{equation}

\noindent The coefficients $t_{ij}$ form a correlation matrix $T_l$ ($l$ = 0,1,2,3 for different projectors). These are given by $T_0$ = diag(-1, -1, -1), $T_1$ = diag(-1, +1, +1), $T_2$ = diag(+1, -1, +1) and $T_3$ = diag(+1, +1, -1). The set of Hadamard projectors on Bob's side is given by $P_x=\frac{1}{2}(I+\mathbf{x}\cdot\sigma)$. Here, $\mathbf{x} = (\pm 1, 0,0)$.

\noindent Now we find the output state of Charlie's qubit, after the two measurements followed by applying appropriate unitaries.
\begin{eqnarray}
    && p_\alpha \varrho_\alpha = \\
    && \mathrm{Tr}_{123} \left[ (P_l \otimes P_x \otimes U_\alpha) (\rho_\S \otimes \rho_{ABC}) (P_l \otimes P_x \otimes U_\alpha^\dagger) \right]. \nonumber
\end{eqnarray}

\noindent The trace is taken over the original qubit, Alice's share, and Bob's share. $\alpha$ acts as multi-index for the pair $(l, x)$. Here, $p_\alpha = \trace{(P_l \otimes P_x \otimes I) (\rho_\S \otimes \rho_{ABC})}$ is the probability of getting the measurement corresponding to the combination $(P_l, P_x)$. Finally, $U_\alpha$ is the unitary operator chosen to reconstruct (a close approximation of) the state at Charlie's side. Substituting the expressions for the states and the projection operators, we obtain:

\begin{eqnarray}
    && \varrho_\alpha = \frac{1}{16 p_\alpha} \Bigg( \bigg[ 1 + \frac{1}{2} \sum_i (T_l)_{ii} A_i \phi_i + \sum_i B_i \phi_i + \nonumber\\
    && \frac{1}{2} \sum_{i,j} (T_l)_{ii} Q_{ij} \phi_i x_j \bigg] I + \sum_{jk} \Omega_{jk} \bigg[ \sum_j C_j + \sum_{ij} S_{ij} x_i \nonumber\\
    && + \frac{1}{2} \sum_{ij} (T_l)_{ii} R_{ij} \phi_i  + \frac{1}{2} \sum_{ijm} (T_l)_{mm} t_{mij} x_i \phi_m \bigg] \sigma_k \Bigg) . 
    \label{eq:final_state}
\end{eqnarray}

\noindent Here, $\{\Omega_\alpha\}$ are rotations in $\mathbb{R}^3$ obtained from the unitaries $\{U_\alpha\}$, given by the relation:

\begin{equation}
    U_\alpha \hat{n}\cdot \sigma U_\alpha^\dagger = (\Omega^\dagger \hat{n}) \cdot \sigma = \sum_{ij} \Omega_{ij}n_i\sigma_j .
\end{equation}

\noindent Now, the expected fidelity of reconstruction, i.e. the "closeness" of Charlie's qubit to the original state, is given by the following integral over the Bloch sphere with uniform distribution $M$:
\begin{equation}
    \mathcal{F} = \oint dM(\phi) \sum_\alpha p_\alpha \trace{\varrho_\alpha \rho_\S} . \label{eq:integral}
\end{equation}

\noindent After substituting expressions from eq~\eqref{eq:final_state} and \eqref{eq:secret}, followed by omitting the terms that do not contribute to the integral, and using the relation,

\begin{equation}
    \oint \langle \phi, Y \phi \rangle dM(\phi) = \frac{1}{3} \trace{Y},
\end{equation}

\noindent the integral in ~\eqref{eq:integral} reduces to

\begin{eqnarray}
    \mathcal{F} = \frac{1}{16} \sum_\alpha \bigg[ 1 + \mathbf{B \cdot x} + \frac{1}{3}\trace{\Omega_\alpha^\dagger R^\dagger T_l} \nonumber\\
    + \frac{1}{3}\trace{\Omega_\alpha^\dagger (\tau_{\lambda\mu\nu}x^\mu)^\dagger T_l}\bigg] .
\end{eqnarray}

\noindent This is being summed up over all $\alpha$, i.e. all the $(l, x)$ possibilities of the two measurements. Note that $\sum_\alpha \mathbf{B \cdot x} = \sum_l \sum_x \mathbf{B \cdot x} = 0$. Let $T$ be the matrix formed by the elements $\{\sum_j t_{ijk}x_j\}$, or in tensor notation,
\begin{equation}
    T = \tau_{\lambda\mu\nu}x^\mu,
    \label{eq:tau_to_t}
\end{equation}
for $\mathbf{x} = (+ 1, 0,0)$. Then, for $\mathbf{x} = (-1, 0,0)$, we have $\tau_{\lambda\mu\nu}x^\mu = -T$. Thus, the summation can be split into two, based on $x$,

\begin{eqnarray}
    \mathcal{F} = \frac{1}{2} + \frac{1}{16}.\frac{1}{3} \sum_l \trace{T_l^\dagger (R+T) \Omega_{(l, +)}} \nonumber\\
    + \frac{1}{16}.\frac{1}{3} \sum_l \trace{T_l^\dagger (R-T) \Omega_{(l, -)}} .
\end{eqnarray}

\noindent Here we have expanded the multi-index $\alpha$ back in the pair form $(l, x)$. Now, we want to choose the rotations to maximise $\mathcal{F}$. As $-T_l^\dagger$ is also a rotation, the $\Omega_\alpha$'s can be chosen independently to maximize each term. We take $\Omega'$ to be the the rotation that maximises the terms corresponding to $T_l^\dagger(R-T)$ and $\Omega''$ for the terms corresponding to $T_l^\dagger(R+T)$. Now, this expression is independent of $l$.

\begin{equation}
    \mathcal{F}_{\max} = \max_{\Omega', \Omega''} \frac{1}{2} \big(1 - \frac{1}{6}\mathrm{Tr} \left[ (R+T)\Omega \right] - \frac{1}{6}\mathrm{Tr} \left[ (R-T)\Omega' \right]\big),
\end{equation}

\noindent where the maximum is taken over all rotations $\Omega', \Omega''$. Since $\Omega'$ and $\Omega''$ can be independent of each other, we get the maximum as 

\begin{eqnarray}
    \mathcal{F}_{\max} = \frac{1}{2} \bigg( 1 + \frac{1}{6} \trace{\sqrt{(R+T)^\dagger(R+T)}} + \nonumber\\
    \frac{1}{6} \trace{\sqrt{(R-T)^\dagger(R-T)}} \bigg). \label{eq:f_max}
\end{eqnarray}

\noindent A tripartite resource state $\varrho$ is useful for reconstruction of the state only when $\mathcal F_{\max} > 2/3$, or when $\vartheta(\varrho) > 1$, where we define $\vartheta(\varrho)$ as:

\begin{equation}
    \vartheta(\varrho) := \frac{1}{2} \bigg( \|R + T\|_1 + \|R - T\|_1 \bigg) , \label{eq:n_rho}
\end{equation}

\noindent such that $\mathcal{F}_{\max} = \dfrac{1}{2}(1 + \dfrac{1}{3}\vartheta(\varrho))$, and $\|\cdot\|_1$ denotes the trace norm of a matrix, given by $\|Z\|_1 = \mathrm{Tr}\sqrt{Z^\dagger Z}$.
Since $R$ shows up in the expression, we conclude that the reconstruction of the state is not entirely because of the controlled reconstruction capability. There can be a situation when $T=O$ ($O$ denotes the null matrix), but the value of $\mathcal{F}$ is greater than $2/3$. Here, as there is no tripartite correlation \cite{42note}, Bob's involvement seems inconsequential. In a sense, this fidelity quantifies the information that can be retrieved as a result of this process. Hence, the CSR fidelity has contributions from both, the reconstruction capacity of the three-qubit resource and the teleportation capacity of the two-qubit channel between the dealer and receiver.

\noindent Here, the dealer-reconstructor subsystem of the resource $\rho_{ABC}$ (from eq~\eqref{eq:resource_state}) is given by:
\begin{align}
    & \rho_{AC} = \mathrm{Tr}_B(\rho_{ABC}) = \\
    & \frac{1}{4}[I^{\otimes 2} +\sum_i a_i.\sigma _i\otimes I + \sum_k I \otimes c_k.\sigma_k+ \sum_{ik} r_{ik}\sigma_i \otimes \sigma_k]. \nonumber
\end{align}
\noindent Using the result from \cite{2}, we can write the teleportation fidelity of $\rho_{AC}$ as:
\begin{equation}
    \mathcal{F}'_{AC} = \frac{1}{2} \big( 1 + \frac{1}{3} \mathrm{Tr} \sqrt{R^\dagger R} \big), \label{eq:teleportation}
\end{equation}

\noindent This analysis is for the case when Alice is the dealer, and the final qubit is being reconstructed at Charlie's end with the assistance of Bob. However, there can be other cases  with the same resource state when the roles are interchanged. This gives us an ordered triplet of (\textbf{dealer, assistant, reconstructor}), which we call the \textit{setting}.

\noindent For some resource states like the GHZ state, all six settings are equivalent due to its symmetry. However, this cannot be generalized, as $\mathcal{F}$ for all the settings need not be the same. We, thus, need to define three different $T$ matrices:
 $T_{AB} = \{Tr(\rho_{ABC}( \sigma_i \otimes \sigma_j \otimes \sigma_x))\}_{ij}$,
 $T_{AC} = \{Tr(\rho_{ABC}( \sigma_i \otimes \sigma_x \otimes \sigma_j))\}_{ij}$ and
 $T_{BC} = \{Tr(\rho_{ABC}( \sigma_x \otimes \sigma_i \otimes \sigma_j))\}_{ij}$.
Here, the subscripts denote the subsystems that contribute to the matrix. For example, in $T_{AB}$, the matrix indices correspond to the first and second subsystems of the three-qubit resource, as seen above. We can, hence, re-write eq~\eqref{eq:n_rho} for the six settings, as shown in table~\ref{tab:settings}.
As the table shows, the expression varies only based on the assistant and is symmetric in swapping the dealer and reconstructor. This symmetry is shared with the expression for fidelity of teleportation \cite{2}.

\begin{table}[ht]
    \centering
    {\renewcommand{\arraystretch}{1.8}
    \begin{tabular}{|c|c|c|}
         \hline \textbf{S. No.} &
         \textbf{Setting} & \textbf{Expression for $\vartheta(\rho)$} \\
         \hline
         1 & (Alice, Bob, Charlie) & $\vartheta_{AC}(\rho) = \displaystyle \frac{1}{2} \big( \|R + T_{AC}\|_1$ \\
         2 & (Charlie, Bob, Alice) & $+ \|R - T_{AC}\|_1 \big)$ \\ 
         \hline
         3 & (Alice, Charlie, Bob) & $\vartheta_{AB}(\rho) = \displaystyle \frac{1}{2} \big(\|Q + T_{AB}\|_1$ \\
         4 & (Bob, Charlie, Alice) & $+ \|Q - T_{AB}\|_1 \big)$ \\ 
         \hline
         5 & (Bob, Alice, Charlie) & $\vartheta_{BC}(\rho) = \displaystyle \frac{1}{2} \big( \|S + T_{BC}\|_1$ \\
         6 & (Charlie, Alice, Bob) & $ + \|S - T_{BC}\|_1 \big)$ \\ 
         \hline
    \end{tabular}}
    \caption{A table showing the expression of $\vartheta(\rho)$ for different settings of (dealer, assistant, reconstructor).}
    \label{tab:settings}
\end{table}

\noindent For simplicity, we will henceforth use the setting (Alice, Bob, Charlie) by default, and follow the representation in eq~\eqref{eq:n_rho}, i.e. $T=T_{AC}$ and $\vartheta(\varrho) = \vartheta_{AC}(\varrho)$, unless specified otherwise.
We have seen that the correlation matrix $S$ is not present in eq~\eqref{eq:n_rho}. It is important to note that this does not rule out the role of Bob in the reconstruction of the state. It only tells that the prior correlation between Bob and Charlie does not affect reconstruction fidelity. The only factors in determining it are the genuine correlation between Alice, Bob and Charlie, captured by $T$, and the correlation matrix between the dealer and the reconstructor, denoted by $R$. 
Hence, we study different cases based on $R, T$ which give an overview of how this score captures both state reconstructing fidelity and the teleportation fidelity of the channel between the source and the reconstructor. These also present us with some conditions on quantum advantage in terms of $R, T$. \\

\noindent \textbf{Case 1: R $\neq$ O, T $\neq$ O} This is the most general scenario when both $R$ and $T$ can take any value. In this case, it is not evident whether the fidelity score is entirely because of the state reconstruction capacity of the entire three-qubit state or, due to the teleportation capacity of the dealer-reconstructor channel, or both. Consequently, we can not pinpoint the main reason behind the quantum advantage. One way to address this situation is to look into the teleportation fidelity of the dealer-reconstructor subsystem. If the teleportation fidelity is at most $2/3$ and the reconstruction fidelity is greater than $2/3$, then we can conjecture that there is a quantum advantage because of the state reconstruction resource. We discuss different cases with the help of the following examples:\\

\noindent \textit{Example 1:} As the first simple case we consider the GHZ state. $\ket{GHZ} = \dfrac{1}{\sqrt 2} (\ket{000}+\ket{111})$. 

\noindent The matrices $R$ and $T$ for the GHZ state can be found after writing the corresponding density state in its Bloch form:
\[ R = \left( \begin{array}{ccc}
0 & 0 & 0\\
0 & 0 & 0\\
0 & 0 & 1
\end{array} \right),
T = \left( \begin{array}{ccc}
1 & 0 & 0\\
0 & -1 & 0\\
0 & 0 & 0
\end{array} \right).
\]
\noindent This gives
\begin{eqnarray}
    \vartheta(\rho_W) = 3, &&
    \mathcal{F}_{\max} = 1 ,
\end{eqnarray}
which is expected since it is already known that the GHZ is used for perfect reconstruction. Note that the teleportation fidelity (from eq~\eqref{eq:teleportation}) for the dealer-reconstructor subsystem, of this state is $\mathcal{F}' = \frac{1}{2}(1 + \frac{1}{3}) = \dfrac{2}{3}$. This clearly gives us a case with a quantum advantage arising from the state reconstruction ability of three-qubit resource states.\\

\noindent \textit{Example 2:}
In a case where both of the fidelities $\mathcal{F}'$ and $\mathcal{F}$ are greater than $2/3$, we cannot be sure whether the quantum advantage in the reconstruction fidelity $\mathcal{F}$ is entirely because of the state reconstruction resource.
The W state, known to exhibit a different nature of entanglement from the GHZ state \cite{43}, is given by $\ket{W} = \frac{1}{\sqrt 3} (\ket{001}+\ket{010}+\ket{100})$. $R$ and $T$ for the W state, from its Bloch representation of $\rho_{W} = \ket{W}\bra{W}$, are:

\[ R = \left( \begin{array}{ccc}
2/3 & 0 & 0\\
0 & 2/3 & 0\\
0 & 0 & -1/3
\end{array} \right),
T = \left( \begin{array}{ccc}
0 & 0 & 2/3\\
0 & 0 & 0\\
2/3 & 0 & 0
\end{array} \right) ,
\]

\noindent which gives
\begin{eqnarray}
    \vartheta(\rho_W) = \frac{7}{3}, &&
    \mathcal{F}_{\max} = \frac{8}{9} \approx 0.89 .
\end{eqnarray}

\noindent We see that $\mathcal{F}_{\max} \neq 1$, which is expected since it is already known that W states cannot be used for perfect controlled reconstruction \cite{44}, but $\vartheta(\rho_W) > 1$ (equivalently, $\mathcal{F}_{\max} > 2/3$). In this case, the subsystem-teleportation fidelity for the dealer-reconstructor channel is found out to be $\mathcal{F}' = \frac{1}{2}(1 + \frac{5}{9}) = \dfrac{7}{9}$. Since in this case $\mathcal{F}'>2/3$, we can not claim that the quantum advantage here is due to genuine tripartite entanglement. \\

\noindent \textit{Example 3:} Next, we consider another example where we show the existence of a state within the paradigm of $R \neq O, T\neq O$, for which the reconstruction fidelity is greater than $2/3$, whereas the teleportation fidelity of the dealer-reconstructor subsystem is $\leq 2/3$. This is a clear example of a state (other than the well-known GHZ state) for which quantum advantage is because of tripartite controlled reconstruction. In this context, let us consider a generalized W Class of states. These states can be expressed as \cite{45, 46}:
\begin{equation}
    \ket{\psi_{W}} = \lambda_0 \ket{000} + \lambda_1 \ket{100} + \lambda_2 \ket{101} + \lambda_3 \ket{110}  \label{eq:acin_w}
\end{equation}

\noindent where $\lambda_i \in \mathbb{R}$, $\lambda_i \geq 0$ and $\sum_i \lambda_i^2 = 1$. For $\rho_{\tilde{W}} = \ket{\psi_{W}}\bra{\psi_{W}}$, the matrices of interest are:
\[ R_{\tilde{W}} = 2 \left( \begin{array}{ccc}
\lambda_0 \lambda_2 & 0 & \lambda_0 \lambda_1\\
0 & - \lambda_0 \lambda_2 & 0 \\
-\lambda_1 \lambda_2 & 0 & \frac{1}{2} - \lambda_1^2 - \lambda_3^2
\end{array} \right),
\]
\[
T_{\tilde{W}} = 2 \left( \begin{array}{ccc}
0 & 0 & \lambda_0 \lambda_3\\
0 & 0 & 0\\
- \lambda_2 \lambda_3 & 0 & -\lambda_1 \lambda3
\end{array} \right).
\]

\begin{figure}[th]
    \includegraphics[width=8cm]{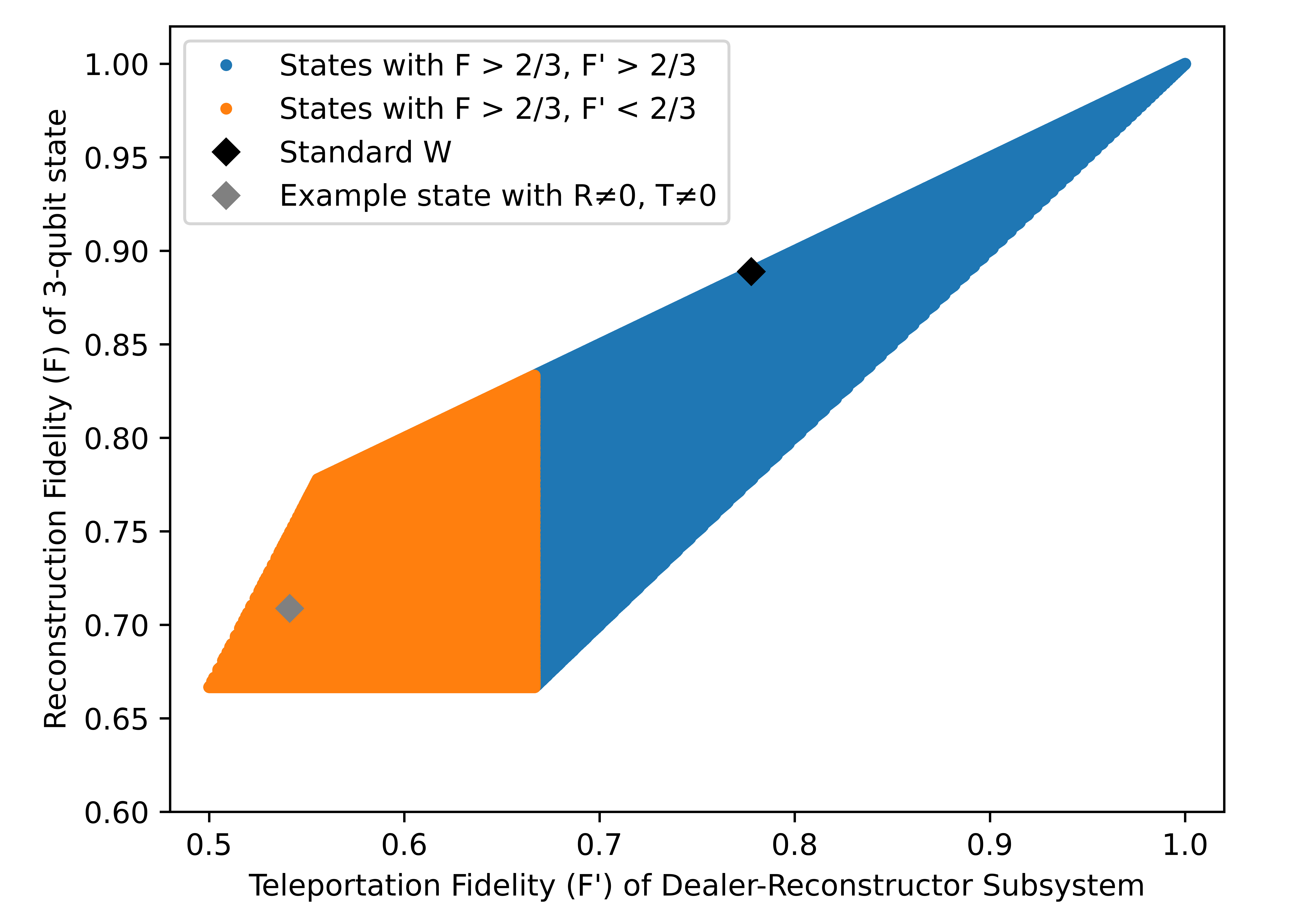}
    \centering
    \caption{Plot comparing the reconstruction fidelity and subsystem-teleportation fidelity for $4\times10^6$ pure state resources uniformly sampled from the entire W class.}
    \label{fig:plot_w}
\end{figure}

In  Fig. \ref{fig:plot_w}, we plot the teleportation fidelity of the dealer-reconstructor subsystem (given by eq~\eqref{eq:teleportation}) against the reconstruction fidelity (given by eq~\eqref{eq:n_rho}) of these states, denoted as $F'$ and $F$ respectively in the figure.
The figure represents the entire set of W states (irrespective of $R,T$), with the orange region depicting states with $F' \le 2/3$, which is of interest, and correspondingly, the blue region has states with $F' > 2/3$. This tells us that states in the orange region are depicting quantum advantage not because of the teleportation channel, while it is still inconclusive for the blue region. Since these are all pure states, $F \ge 2/3$. Not all of these states satisfy $R \neq O, T \neq O$, but one state that falls in this paradigm has been marked in the figure as an example of quantum advantage, as it lies in the orange region. This state is described by parameters $\lambda_0 = \lambda_1 = 0.7, \lambda_2 \approx 0.09, \lambda_3 \approx 0.11$. The standard W state has also been marked, which lies in the blue region, as expected from the discussion in Example 2.\\

\noindent \textbf{Case 2: R = O, T $\neq$ O} In this case, since the correlation matrix $R = O$, it can be said with certainty that there is no direct correlation between the dealer and the reconstructor. This means there will be no teleportation capacity of the channel between them, and hence, it will not affect the total reconstruction fidelity. In other words, if there is a quantum advantage in this region, we can surely claim that this is because of tripartite's controlled reconstruction capacity. However, the converse is not true. There can be states with $R \neq O$ for which there is no correlation between the dealer and the reconstructor. In other words, there can be quantum states whose quantum advantage arises solely from the $T$ component of correlation even when $R \neq O$. \\

\noindent \textbf{Theorem 1 :} If R = O, T $\neq$ O, the observed quantum advantage is only because of tripartite state reconstruction.\\

\noindent \textbf{Proof:}
\noindent Putting $R=O$ in the expression for subsystem-teleportation fidelity (from eq~\eqref{eq:teleportation}), gives $\mathcal{F}' = \dfrac{1}{2}$. Hence, no quantum information flow can flow from dealer to reconstructor through teleportation alone. Then, the contribution to the reconstruction fidelity has to come from the tripartite channel. $\blacksquare$\\

\noindent \textit{Example:} Consider the following states:
$$\ket{\gamma_\pm} = \frac{ \ket{000} \pm \ket{100} \pm \ket{110} + \ket{111} }{2}, $$
\noindent Both these states fall into Case 1 but if we take their equal mixture
$\rho_\gamma = \frac{1}{2} (\ket{\gamma_-}\bra{\gamma_-} + \ket{\gamma_+}\bra{\gamma_+}),$
\noindent  $R$ is found to be $O$. Moreover, we get $\mathcal{F} = 3/4 > 2/3$, giving us a quantum advantage. Since $R = O$ and teleportation cannot contribute to the final fidelity, we infer that this advantage arises purely from the tripartite state reconstruction. \\

\noindent \textbf{Case 3: R $\neq$ O, T = O} If in equation ~\eqref{eq:n_rho}, $R \neq O$, but $T=O$, then it can be said that there is no tripartite correlation and hence there is no involvement of Bob. So, in principle, there is no question of controlled reconstruction in this case. In such a case, if the reconstruction fidelity is greater than $2/3$, that is purely because of the teleportation capacity of the subsystem. However, the converse is not true as there can be tripartite states with $T \neq O$ but not having genuine quantum correlation.\\

\noindent \textbf{Theorem 2 :} If $R\neq O$ and $T=O$, then the quantum advantage in the reconstruction is entirely because of the teleportation capacity of the subsystem between the dealer and the reconstructor.\\

\noindent{\textbf{Proof :}} Putting $T=O$ in eq~\eqref{eq:n_rho}, we get the expression
\begin{equation}
    \vartheta(\rho_{ABC}) = \|R\|_1.
\end{equation}

\noindent The expression in eq~\eqref{eq:teleportation} now matches with the expression for reconstruction fidelity given by
\begin{equation*}
    \mathcal{F}_{\max} = \dfrac{1}{2} (1 + \dfrac{1}{3}\vartheta(\rho_{ABC})) = \dfrac{1}{2} (1 + \dfrac{1}{3}\|R\|_1).
    \end{equation*}
Hence, it can be concluded that quantum advantage in this case is solely because of the teleportation channel between the dealer and the reconstructor. $\blacksquare$\\

\noindent \textit{Example:} Consider the following states:

$$\ket{\delta_\pm} = \frac{ \ket{000} \pm \ket{100} \pm \ket{110} + \ket{111} }{2}, $$

\noindent Both of these have $R \neq O \neq T$, but when their equal mixture is considered, then for the mixed state $\rho_\delta$, we find $T = O$, where $\rho_\delta = \frac{1}{2} (\ket{\delta_-}\bra{\delta_-} + \ket{\delta_+}\bra{\delta_+}),$
\noindent In this case as well, we get $\mathcal{F} = 3/4 > 2/3$. However, since $T$ is $O$, we argue that this fidelity arises solely from the teleportation capacity of the channel between the sender and the reconstructor and, hence, the quantum advantage. \\

\noindent \textbf{Case 4: R = O, T = O} Here, $\vartheta(\rho) = 0$, giving us $\mathcal{F}_{\max} = 1/2$ which is no better than a random guess. In this case, there cannot be any quantum advantage because there is no flow of information from the dealer to the reconstructor.\\

\section{Quantum Secret Sharing} \label{sec:secret}

For any secret-sharing protocol to be successful, neither of the shareholders should have useful information about the secret on their own. In other words, Charlie should not be able to reconstruct the secret without Bob's involvement, and vice-versa. In this context, we use the term \textit{useful information} as the amount of extra information that can be obtained over the classical limit of the respective channels. One quantification can be the teleportation capability of the bipartite channels between the dealer and the shareholders. If this is more than $2/3$, then there is an information gain through the bipartite channel  compared to what can be achieved classically.  This enforces additional conditions on the resource to ensure the protocol is secure against dishonest parties. The maximum expected fidelity that Bob can obtain on his own accord is equivalent to the teleportation capacity of subsystem $\rho_{AB}$ of the resource since Charlie is not involved. The state $\rho_{AB}$ is given by,
\begin{align*}
    & \rho_{AB} = \mathrm{Tr}_C(\rho_{ABC}) = \\
    & \frac{1}{4}[I^{\otimes 2} +\sum_i a_i.\sigma _i\otimes I + \sum_j I \otimes b_j.\sigma_j+ \sum_{ij} q_{ij}\sigma_i \otimes \sigma_j].
\end{align*}
The teleportation capacity of this bipartite resource, i.e. the maximum expected share of Bob without involving Charlie, is:
\begin{equation}
    \mathcal{F}'_{AB} = \frac{1}{2} \big( 1 + \frac{1}{3} \mathrm{Tr} \sqrt{Q^\dagger Q} \big), \label{eq:tp_bob}
\end{equation}

\noindent We have already seen the expression of $\mathcal{F}'_{AC}$ for Charlie's case in eq~\eqref{eq:teleportation}. Since we know that a state is useful for quantum teleportation when $\mathcal{F}' > 2/3$ \cite{2}, we want:
\begin{eqnarray}
    \mathcal{F}'_{AB} \le \frac{2}{3} && \mathcal{F}'_{AC} \le \frac{2}{3} \nonumber \\
    \implies \mathrm{Tr} \sqrt{Q^\dagger Q} \le 1 && \mathrm{Tr} \sqrt{R^\dagger R} \le 1 \label{eq:extra_conditions}
\end{eqnarray}

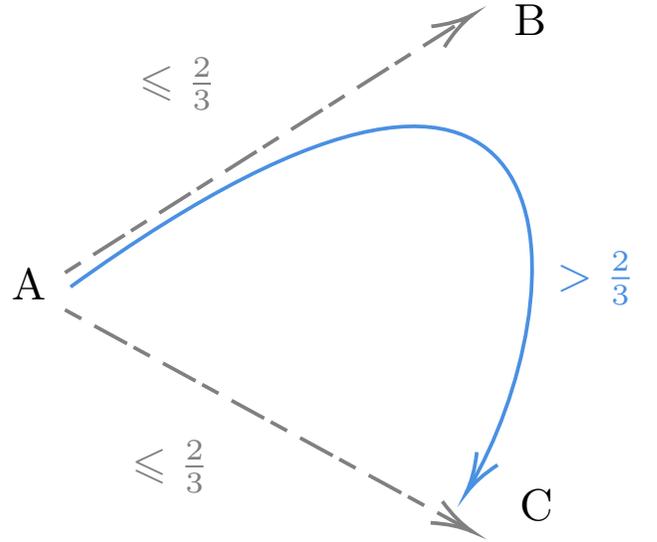
\begin{figure}[hbt]
    \centering
    \resizebox{\columnwidth}{!}{\input{fig2}}
    \caption{Conditions for secret sharing \linebreak
    (A: dealer, B: assistant, C: reconstructor)}  \label{fig:tikz}
\end{figure}

\noindent These constraints, along with $\vartheta(\rho) > 1$ (or, equivalently, $\mathcal{F} > 2/3$), together form the three conditions for successful QSS (see Fig. \ref{fig:tikz}).  This ensures that the secret is faithfully reconstructed.
Additionally it accounts for dishonest recipients (either Bob or Charlie) as this ensures that the information they can retrieve from the quantum resource, without faithfully cooperating does not exceed the information that can be obtained  classically even in absence of quantum channel. Thus, we are able to provide a way to characterize the states useful for secret sharing more precisely.\\

\noindent \textit{Example:} Consider the following states:

$$\ket{\beta_\pm} = \frac{ \ket{000} + \ket{100} + \ket{101} \pm \ket{110} }{2}, $$

\noindent The equal mixture of $\ket{\beta_+}\bra{\beta_+}$ and $\ket{\beta_-}\bra{\beta_-}$ gives  $\mathrm{Tr} \sqrt{Q^\dagger Q} = 1/2 = \mathrm{Tr} \sqrt{R^\dagger R}$, and still gives $\mathcal{F} > 2/3$, fulfilling all three conditions. This example demonstrates successful secret sharing, in addition to the well-known GHZ state. This is contrary to the belief that only the GHZ state is useful for secret sharing.  \\

\section{Discussion}
In this article, we have given an expression for the maximum expected CSR fidelity for general three-qubit states in terms of their Bloch parameters. This fidelity has contributions from both the tripartite correlation tensor and the correlation matrix between the dealer-reconstructor subsystem. We report a quantum advantage in the reconstruction of the state over its classical limit ($\frac{2}{3}$). Different cases, in terms of the involved correlation matrices, are then investigated with respect to the CSR fidelity. We provide examples where the advantage is only because of the teleportation capacity of the subsystem and those where the advantage is mainly because of the state reconstruction capacity of the tripartite resource. We also discuss cases which are ambiguous when it comes to the cause of the quantum advantage. Our results introduce interoperability between quantum teleportation and CSR. This also opens up new avenues of research related to the interoperability that can exist in different quantum information processing tasks.
We extend our analysis by outlining additional conditions that must be satisfied to ensure the protocol is secure by preventing any dishonest participant from having a quantum advantage. These conditions distinguish CSR from QSS in a true sense. Further work can be done in this context, analysing the interference of eavesdroppers, and how they can be detected. From a resource point of view, we are able to characterize the states that provide quantum advantage in CSR. In addition, we are able to identify the states that can be used as resources for QSS.\\

\section{Acknowledgements}
The authors thank Dr Shantanav Chakrabarty (Centre for Quantum Science \& Technology, IIIT Hyderabad) and Dr Nirman Ganguly (Department of Mathematics, BITS-Pilani Hyderabad) for their valuable inputs.



\appendix
\section{The Case of Dishonest Party in Classical Scenario \label{ss:dishonest}}

\noindent Here, we wish to see what happens in the case if one of the receiving parties (either Bob or Charlie) is dishonest. Since information from both the parties is required to reconstruct the qubit, the honest one can keep a check on the dishonest one. But we also have to take a look at what might happen in case the dishonest person wants to guess the shared qubit based on what information they have.

\noindent Before we move on, let us have a look at the distributions of the values $s_1, s_2$ can take. Let $s_i \leftarrow \{p_0, p_1\}$ denote that $s_i$ takes the value 0 with probability $p_0$ and 1 with probability $p_1$. Now if we have $s_1 \leftarrow \{p, 1-p\}$, then either $s_2 \leftarrow \{p, 1-p\}$ (if $s=0$) or $s_2 \leftarrow \{1-p, p\}$ (if $s=1$). In either case, both the distributions are dependant on each other, and once one of the bits is sampled, the other bit is known with certainty.

\noindent Since the distributions are symmetric, without the loss of generality, we can take Bob* as the dishonest party. He knows $s_1$ but has no knowledge of $s_2$ without communicating with Charlie. So he has to guess the bit $s$, and let's denote his guess of the bit as $s'$ which is either same as his bit ($s_1$) or the negation of it.

\noindent \textbf{Case - 1} $s' = s_1$, or equivalently, Bob* guesses $s_2 = 0$

\noindent Let $s_2 \leftarrow \{p, 1-p\}$. The fidelity between the original qubit and the one reconstructed by Bob can be expressed as: ($F'$ is used to denote fidelity with guess).

\begin{widetext}
    \begin{align*}
        F'(q, s') & = Pr[s = s']F(q, s) + Pr[s \neq s']F(q, \neg s)                                                                                                                                                                                                                           \\
                  & = Pr[s_2 = 0] \big( Pr[s = 0] \cdot \| \braket{q|\uparrow} \|^2 + Pr[s = 1] \cdot \| \braket{q|\downarrow} \|^2 \big)                                                                                                                                                     \\ & \hspace{40pt} + Pr[s_2 = 1] \big( Pr[\neg s = 0] \cdot \| \braket{q|\uparrow} \|^2 + Pr[\neg s = 1] \cdot \| \braket{q|\downarrow} \|^2 \big) \\
                  & = (p) \big( \cos^2{\frac{\theta}{2}} \cdot \cos^2{\frac{\theta}{2}} + \sin^2{\frac{\theta}{2}} \cdot \sin^2{\frac{\theta}{2}} \big) + (1-p) \big( \sin^2{\frac{\theta}{2}} \cdot \cos^2{\frac{\theta}{2}} + \cos^2{\frac{\theta}{2}} \cdot \sin^2{\frac{\theta}{2}} \big) \\
                  & = p \big( 1 - \frac{1}{2} \sin^2\theta \big) + \frac{1}{2} (1 - p) \sin^2\theta = p \cdot \cos^2\theta + \frac{1}{2}\sin^2\theta
    \end{align*}
\end{widetext}

\noindent Hence the expected "guess" fidelity in this case:

\begin{widetext}
    \begin{align*}
        \mathcal{F'} = \langle F'(q, s') \rangle & = \frac{1}{4\pi} \int_{\theta = 0}^\pi \int_{\phi = 0}^{2\pi} \sin\theta \,d\theta \,d\phi (p \cdot \cos^2\theta + \frac{1}{2}\sin^2\theta) \\
                                                 & =  \frac{1}{4\pi} \cdot 2\pi \Big( \frac{4}{3} - \frac{2}{3}p \Big) = \boxed{\frac{2 - p}{3}}
    \end{align*}
\end{widetext}

\noindent We want Bob*'s guess to be no better than a random one. Or equivalently, we want

\begin{equation}
    \mathcal{F'} \leq \frac{1}{2} \implies \frac{2 - p}{3} \leq \frac{1}{2} \implies \boxed{p \geq \frac{1}{2}}
\end{equation}

\noindent \textbf{Case - 2} $s' = \neg s_1$, or equivalently, Bob* guesses $s_2 = 1$

\noindent In this case,

\begin{align*}
    F'(q, s')    & = Pr[s_2 = 1] F(q, s) + Pr[s_2 = 0] F(q, \neg s)      \\
                 & = (1-p) \cos^2\theta + \frac{1}{2}\sin^2\theta        \\
    \intertext{and}
    \mathcal{F'} & = \langle F'(q, s') \rangle = \boxed{\frac{1 + p}{3}}
\end{align*}

\noindent Enforcing the same condition as last time,

\begin{equation}
    \mathcal{F'} \geq \frac{1}{2} \implies \frac{1 + p}{3} \leq \frac{1}{2} \implies \boxed{p \leq \frac{1}{2}}
\end{equation}

\noindent Both the conditions will be satisfied if we set $p = \frac{1}{2}$. Hence, Alice can prevent either party from cheating if the uniform distribution is used to sample one of the bits.

\end{document}

%% file: fig2.tex
\begin{tikzpicture}[x=0.75pt,y=0.75pt,yscale=-1,xscale=1]

\draw [color={rgb, 255:red, 128; green, 128; blue, 128 }  ,draw opacity=1 ] [dash pattern={on 3.75pt off 3pt on 7.5pt off 1.5pt}]  (27.67,74.67) -- (135.48,6.4) ;
\draw [shift={(137.17,5.33)}, rotate = 147.66] [color={rgb, 255:red, 128; green, 128; blue, 128 }  ,draw opacity=1 ][line width=0.75]    (10.93,-3.29) .. controls (6.95,-1.4) and (3.31,-0.3) .. (0,0) .. controls (3.31,0.3) and (6.95,1.4) .. (10.93,3.29)   ;
\draw [color={rgb, 255:red, 128; green, 128; blue, 128 }  ,draw opacity=1 ] [dash pattern={on 3.75pt off 3pt on 7.5pt off 1.5pt}]  (27.67,84.67) -- (135.41,143.38) ;
\draw [shift={(137.17,144.33)}, rotate = 208.59] [color={rgb, 255:red, 128; green, 128; blue, 128 }  ,draw opacity=1 ][line width=0.75]    (10.93,-3.29) .. controls (6.95,-1.4) and (3.31,-0.3) .. (0,0) .. controls (3.31,0.3) and (6.95,1.4) .. (10.93,3.29)   ;
\draw [color={rgb, 255:red, 74; green, 144; blue, 226 }  ,draw opacity=1 ]   (29.17,78.5) .. controls (171.73,-24.63) and (165.64,82.81) .. (136.39,132.51) ;
\draw [shift={(135.5,134)}, rotate = 301.48] [color={rgb, 255:red, 74; green, 144; blue, 226 }  ,draw opacity=1 ][line width=0.75]    (10.93,-3.29) .. controls (6.95,-1.4) and (3.31,-0.3) .. (0,0) .. controls (3.31,0.3) and (6.95,1.4) .. (10.93,3.29)   ;

\draw (11.83,72.17) node [anchor=north west][inner sep=0.75pt]   [align=left] {A};
\draw (146.83,0.83) node [anchor=north west][inner sep=0.75pt]   [align=left] {B};
\draw (148.5,131.5) node [anchor=north west][inner sep=0.75pt]   [align=left] {C};


\draw (46.17,15.23) node [anchor=north west][inner sep=0.75pt]  [color={rgb, 255:red, 128; green, 128; blue, 128 }  ,opacity=1 ]  {$\leqslant \frac{2}{3}$};
\draw (44.17,118.23) node [anchor=north west][inner sep=0.75pt]  [color={rgb, 255:red, 128; green, 128; blue, 128 }  ,opacity=1 ]  {$\leqslant \frac{2}{3}$};
\draw (158.5,67.57) node [anchor=north west][inner sep=0.75pt]  [color={rgb, 255:red, 74; green, 144; blue, 226 }  ,opacity=1 ]  {$ >\frac{2}{3}$};

\end{tikzpicture}

%% file: main.bbl
\begin{thebibliography}{99}
    \bibliographystyle{unsrt}

\bibitem{1} C. H. Bennett, G. Brassard, C. Crepeau, R. Jozsa, A. Peres and W. K. Wootters, Phys. Rev. Lett. \textbf{70}, 1895 (1993)
\bibitem{2} R. Horodecki, M. Horodecki, P. Horodecki, Phys.Lett. A \textbf{222}, 21 (1996).
\bibitem{3} I. Chakrabarty, Eur. Phys. J. D \textbf{57}, 265-269 (2010).
\bibitem{4} S. Adhikari, N. Ganguly, I. Chakrabarty, B. S. Choudhury, J. Phys. A: Math. Theor. \textbf{41}, 415302 (2008).
\bibitem{5} C. H. Bennett, P. Hayden, D. W. Leung, P. W. Shor, and Andreas Winter, IEEE Transactions on Information Theory, Vol. \textbf{51}, Part 1 (2005).
\bibitem{6}Sohail, A. K Pati, V. Aradhya, I. Chakrabarty, S. Patro, arXiv:2302.11499.
\bibitem{7} A. K. Pati Phys. Rev. A \textbf{63}, 014302.

\bibitem{8}  C. H. Bennett and S. J. Wiesner, Phys. Rev. Lett. \textbf{69}, 2881 (1992)
\bibitem{9} C. Srivastava, A. Bera, A. Sen De, U. Sen, Phys. Rev. A \textbf{100}, 052304 (2019).
\bibitem{10} T. Das, R. Prabhu, A. Sen De, U. Sen; Phys. Rev. A \textbf{92}, 052330 (2015).
\bibitem{11} S. Patro, I. Chakrabarty, N. Ganguly, Phys. Rev. A \textbf{96}, 062102 (2017).
\bibitem{12} M. Vempati, N. Ganguly, I. Chakrabarty, A. K. Pati, Phys. Rev. A \textbf{104}, 012417 (2021).
\bibitem{13} S. Roy, T. Chanda, T. Das, A. Sen De, U. Sen,  Phys. Lett. A \textbf{382}, 1709 (2018).

\bibitem{14} S. Wehner, D. Elkouss, and R. Hanson, Science \textbf{362} (2018).
\bibitem{15} J. Biamonte, M. Faccin, and M. De Domenico,  Commn. Phys. \textbf{2}, 1 (2019).
\bibitem{16} G. Chiribella, G. M. D’Ariano, and P. Perinotti, Phys. Rev. A \textbf{80}, 022339 (2009).
\bibitem{17} H Jeff Kimble,  Nature \textbf{453}, 1023 (2008).
\bibitem{18} C. Simon, Nature Photonics \textbf{11}, 678 (2017).
\bibitem{19} Sk. Sazim and I. Chakrabarty,  Eur. Phys. J D, \textbf{67}, 174 (2013).
\bibitem{20} K. Mukherjee, I. Chakrabarty, G. Mylavarapu, Phys. Rev. A \textbf{107}, 032404 (2023).
\bibitem{21} M. K. Shukla, M. Huang, I. Chakrabarty, J. Wu, arXiv:2011.07554.


\bibitem{22} M. Hillery, V. Buzek, and A. Berthiaume, Phys.
Rev. A \textbf{59}, 1829 (1999).
\bibitem{23} R. Cleve, D. Gottesman, H.K. Lo, Phys.Rev.Lett. \textbf{83}  648 (1999).
\bibitem{24}A. Karlsson, M. Koashi, and N. Imoto, Phys. Rev. A \textbf{59}, 162 (1999).
\bibitem{25} S. Bandyopadhyay, Phys. Rev. A \textbf{62}, 012308
(2000).
\bibitem{26} D. Markham, B.C. Sanders, Phys. Rev. A \textbf{78}, 042309 (2008).
\bibitem{27} Q. Li, W. H. Chan, and D-Y Long, Phys. Rev. A \textbf{82}, 022303 (2010).
\bibitem{28} S.Adhikari, I. Chakrabarty, P. Agrawal, QIC, \textbf{12}, 0253 (2012).
\bibitem{29} M. Ray, S. Chatterjee, and I. Chakrabarty,  Eur. Phys. J. D \textbf{70}, 114 (2016).
\bibitem{30} S. Adhikari, Quantum secret sharing with two qubit bipartite mixed states, arXiv:1011.2868.
\bibitem{31} S. Sazim, C. Vanarasa, I. Chakrabarty, K. Srinathan, Quant. Inf. Proc. \textbf{14}, 12, 4651 (2015).
\bibitem{32} W. Tittel, H. Zbinden, and N. Gisin, Phys. Rev. A \textbf{63}, 042301 (2001).
\bibitem{33} C. Schmid, P. Trojek, M. Bourennane, C. Kurtsiefer, M. Zukowski, and H. Weinfurter, Phys. Rev. Lett. \textbf{95}, 230505 (2005).
\bibitem{34} C. Schmid, P. Trojek, S. Gaertner, M. Bourennane, C. Kurtsiefer, M. Zukowski, and H. Weinfurter, Fortschritte der Physik \textbf{54}, 831 (2006).
\bibitem{35} J. Bogdanski, N. Rafiei, and M. Bourennane, Phys. Rev. A \textbf{78}, 062307 (2008).
\bibitem{36}  P. Mudholkar, C. Vanarasa, I. Chakrabarty, S. Kannan, arXiv:2112.15556 .

\bibitem{37} X. H. Li and S. Ghose, Phys. Rev. A \textbf{90}, 052305 (2014)
\bibitem{38} A. Kumar, S. Haddadi, M.R. Pourkarimi, et al., Sci Rep \textbf{10}, 13608 (2020)
\bibitem{39} S. Gangopadhyay, T. Wang, A. Mashatan, and S. Ghose, Phys. Rev. A \textbf{106}, 052433 (2022).
\bibitem{40} A. Garg and S. Adhikari, Estimation of Power in the Controlled Quantum Teleportation through the Witness Operator, arXiv:2307.16574.

\bibitem{41} A. Karlsson and M. Bourennane, Phys. Rev. A \textbf{58}, 4394 (1998).

\bibitem{42note} In the later sections of this paper, the term tripartite correlation/correlation tensor is sometimes used to talk about the matrix $T$. While it is important to note that although it is derived from the tensor $\tau$ of order 3, $T$ itself has order 2, and hence is described by a matrix (The reader can refer to eq~\eqref{eq:tau_to_t} for clarification). Though we can use this terminology for simplicity, since $T$ does capture a part of the tripartite correlation.

\bibitem{43} W. Dur, G. Vidal, and J. I. Cirac, Phys. Rev. A \textbf{62}, 062314 (2000).
\bibitem{44} A. Das, S. Nandi, Sk. Sazim, P. Agrawal, Eur. Phys. J. D \textbf{74}, 91 (2020).
\bibitem{45} A. Acin, D. Bruss, M. Lewenstein, and A. Sanpera, Phys. Rev. Lett. \textbf{87}, 040401 (2001).
\bibitem{46} P. Pandya, A. Misra, I. Chakrabarty,  Phys. Rev. A \textbf{94}, 052126 (2016).
\end{thebibliography}
